\documentclass[prb,
twocolumn,
superscriptaddress,showpacs,amsmath,amssymb]{revtex4}
\usepackage{color}
\usepackage{soul}

\begin{document}

\title{Vielbein with mixed dimensions and gravitational global monopole in the planar phase of superfluid $^3$He}

\author{G.E.~Volovik}
\affiliation{Low Temperature Laboratory, Aalto University,  P.O. Box 15100, FI-00076 Aalto, Finland}
\affiliation{Landau Institute for Theoretical Physics, acad. Semyonov av., 1a, 142432,
Chernogolovka, Russia}

\date{\today}

\begin{abstract}
The planar phase of superfluid $^3$He has Dirac points in momentum space and the analog of Dirac monopole in the real space. 
Here we discuss the combined effect of Dirac point and Dirac monopole. It is shown that in the presence of the monopole the effective metric acting on Dirac fermions corresponds to the metric produced by the global monopole in general relativity: it is the conical metric.
Another consequence is that the primary variable, which gives rise to the effective metric, is the unusual vielbein field in the form of the $4\times 5$ matrix, as distinct from the conventional $4\times 4$ matrix of the tetrad field in tetrad gravity. 
\end{abstract}

\maketitle

\section{Introduction}

The planar phase is one of the possible superfluid phases of liquid $^3$He.\cite{Vollhardt1990}
 It may exist in some region of the phase diagram of superfluid $^3$He confined in aerogels.\cite{Surovtsev2019} 
The planar phase has two Dirac points in the quasiparticle spectrum, which are supported by combined action of topology and some special symmetry, see e.g. Ref. \cite{Makhlin2014}. The quasiparticles in the planar phase with fixed spin behave as Weyl fermions. Similar to the chiral superfluid $^3$He-A, they experience the effective gravity and gauge field produced by the deformation of the order parameter. But there is the following important difference. In $^3$He-A, the spin-up and spin-down fermions have the same chirality, while in the planar phase the spin-up and spin-down fermions have the opposite chirality. As a result the Weyl fermions in planar phase form the massless Dirac fermions, see Ref. \cite{Volovik2003}. 

Here we study the planar phase fermions in the presence of the topological defect -- the hedgehog. The effective gravity produced by the hedgehog appears to be similar to the gravitational effect of the global monopole in general relativity: it gives rise to the conical space.\cite{Starobinskii1977,Barriola1989,Monopole1990,Monopole2001,Bronnikov2002,Mavromatos2017,Petrov2019,Monopole2020}
Another consequence of the hedgehog is that the vielbein, which describes the effective gravity, is the $4\times 5$ matrix, as distinct from the conventional $4\times 4$ matrix in the tetrad formalism of general relativity.

\section{Weyl-Dirac points and $4\times 5$ vielbein}

In the general spin triplet $p$-wave pairing state the symmetric $2\times 2$ matrix of the gap function is:
\begin{equation}
 \hat{\Delta}=  A_{\alpha}^i\sigma^\alpha p_i \,,
\label{SpinTriplet}
\end{equation}
where $\sigma^\alpha$ are the Pauli matrices for spin and $A_{\alpha i}$  is the $3\times 3$ complex matrix of the order parameter, see the book \cite{Vollhardt1990}.

In  the planar phase the particular representative of the order parameter is:
\begin{eqnarray}
A_{\alpha i}=c_\perp e^{i\Phi}\left(\delta_\alpha^i - {\hat l}_\alpha  {\hat l}^i \right) \,,
\label{OP1}
\end{eqnarray}
where $\Phi$ is the phase of the order parameter and $\hat l$ is the unit vector. 
All the other degenerate states of the planar phase are obtained by spin, orbital and phase rotations of the group $G= SO(3)_S \times  SO(3)_L \times U(1)$
(here we ignore the discrete symmetry, since we are only interested in the global monopole).

The order parameter in Eq.(\ref{OP1}) has the symmetry $H=SO(2)_J$ -- the symmetry under the common spin and orbital rotations about the axis $\hat l$. As the result, the manifold of the degenerate states is $R=(SO(3)_S \times  SO(3)_L \times U(1))/SO(2)_J$, which supports the 
monopoles (hedgehogs), described by the homotopy group $\pi_2(R)=Z$. 
The particular form of the monopole with the topological charge $N=1$ is:
\begin{eqnarray}
A_{\alpha i}({\bf r})=f(r)\left(\delta_\alpha^i - {\hat r}_\alpha  {\hat r}^i \right) \,,
\label{Monopole}
\end{eqnarray}
where $ {\hat r}= {\bf r}/r$,  and  $f(r\rightarrow \infty)=c_\perp$. We also fix the phase $\Phi=0$ in further considerations.

The Bogoliubov-Nambu Hamiltonian for quasiparticles:
\begin{equation}
  \begin{pmatrix}\epsilon(p) &\hat \Delta
  \\
  \hat\Delta^\dagger& -\epsilon(p)
  \end{pmatrix}
   \,,
\label{HamiltonianGeneral}
\end{equation}
where $\epsilon(p)=c_\parallel(p-p_F)$, $c_\parallel=v_F$, and $v_F$ and $p_F$ are correspondingly the Fermi velocity and Fermi momentum of the normal Fermi liquid.

The planar phase has the Weyl-Dirac points at ${\bf p}=\pm p_F\hat l$. Near the Weyl-Dirac nodes the Hamiltonian has the form:
\begin{equation}
  H=\sum_a \Gamma^a e_a^i (p_i - qA_i)  \,.
\label{Hamiltonian}
\end{equation}
 Here ${\bf A}=p_F\hat l$ is the vector potential of effective gauge field acting on the massless Dirac fermions; $q=\pm 1$ is the corresponding electric charge; $\Gamma^a$ with $a=1,2,3,4$ are the gamma matrices; and $e_a^i$ are the components of the spatial vielbein  with $a=1,2,3,4$ and $i=1,2,3$.
 
 The Hamiltonian (\ref{Hamiltonian}) contains four $4\times 4$ Hermitian $\Gamma$-matrices with $\{\Gamma^a,\Gamma^b\}= 2\delta^{ab}$, which here we choose in the following way:
\begin{equation}
\Gamma^1=\tau_1\sigma_x\,\,, \, \Gamma^2=\tau_1\sigma_y\,\,, \,    \Gamma^3=\tau_1\sigma_z \,\,,\,  \Gamma^4=\tau_3  \,,
\label{Gamma}
\end{equation}
 and the corresponding vielbein components are
 \begin{equation}
 e^i_a= c_\perp (\delta_a^i -\hat l_a \hat l^i )\,\,\,  {\rm for} \,\, a=1,2,3\,\,, \, e^i_4=c_\parallel \hat l^i\,.
\label{vielbein}
\end{equation}

 The important property of such vielbein is that it is the $3\times 4$ matrix, instead of the conventional $3\times 3$ matrix of the dreibein. Nevertheless,  this asymmetric vielbein provides the correct expression for the elements of the effective metric:
  \begin{equation}
g^{ik}=\sum_{a,b}\delta^{ab} e^i_a  e^k_b \,\,, \,  a,b=1,2,3,4  \,\,, \,  i,k=1,2,3 \,,
\label{3Dmetric1}
\end{equation}
This metric has the conventional form
  \begin{equation}
g^{ik}= c_\parallel^2 \hat l^i  \hat l^k + c_\perp^2 (\delta^{ik}-\hat l^i  \hat l^k)\,,
\label{3Dmetric2}
\end{equation}
which coincides with the effective metric in the chiral superfluid $^3$He-A with Weyl nodes, see Ref.\cite{Volovik2003}.

The $3+1$ effective metric is expressed in terms of the $4\times 5$ vielbein,
 \begin{equation}
g^{\mu\nu}=\sum_{a,b}\eta^{ab} e^\mu_a  e^\nu_b \,\,, \, a,b=0,1,2,3,4 \,\,, \, \mu,\nu=0,1,2,3\,.
\label{4Dmetric1}
\end{equation}
It also has the form as in $^3$He-A. 
In spite of the unusual asymmetric $4\times 5$ vielbein, which is not invertible, the effective metric is well defined and is invertible:
 \begin{eqnarray}
g_{ik}= \frac{1}{c_\parallel^2} \hat l_i  \hat l_k + \frac{1}{c_\perp^2} (\delta_{ik}-\hat l_i  \hat l_k) \,\, , \, g_{00}=-1\,.
\label{InverseMetric}
\end{eqnarray}

\section{Global monopole and effective metric}

For the monopole, the asymptotic form of the metric at infinity, where $f(r)\rightarrow c_\perp$, is:
 \begin{equation}
g^{ik}({\bf r}) =c_\perp^2 \delta^{ik} + (c_\parallel^2 - c_\perp^2){\hat r}^i {\hat r}^k\,,
\label{MetricMonopole}
\end{equation}
and the interval is:
 \begin{equation}
ds^2= -dt^2 + \frac{1}{c_\perp^2} r^2 d\Omega^2  +\frac{1}{c_\parallel^2} dr^2\,.
\label{ds}
\end{equation}
Eq.(\ref{ds})  represents conical spacetime, which in general relativity is  produced by the global monopoles (monopoles without gauge fields, see Refs. \cite{Starobinskii1977,Barriola1989,Monopole1990,Monopole2001,Bronnikov2002,Mavromatos2017,Petrov2019,Monopole2020}). This spacetime has the nonzero scalar curvature:
 \begin{equation}
R=2\frac{1-\alpha^2}{r^2} \,\,,\, \alpha^2= \frac{c_\parallel^2}{c_\perp^2} \,.
\label{curvature}
\end{equation}
The analog of the global monopole was considered in superfluid $^3$He-A (see Refs. \cite{Volovik1998,Monopole2001}). However, in $^3$He-A this hedgehog in the $\hat l$ field has the tail -- the doubly quantized vortex. This is  the analog of the Nambu monopole  \cite{Nambu1977} terminating cosmic string -- the observable Dirac string (classification of such composite objects can be found in Ref. \cite{Zhang2020}). 
In the planar phase the monopole is topologically stable and the Nambu-Dirac string does not appear. This is because the Dirac string of the  monopole in the orbital vector $\hat l^i ({\bf r})=\hat r^i $ in Eq.(\ref{Monopole}) is cancelled by the Dirac string from the monopole in the spin vector $\hat l_\alpha({\bf r}) = \hat r_\alpha $.

Since in superfluid $^3$He one has $c_\parallel^2 >c_\perp^2$,  the effective metric corresponds to the spacetime with the solid angle excess,\cite{Volovik1998,Monopole2001}  $\alpha^2>1$, instead of the solid angle deficit discussed for the global cosmic monopoles with  $\alpha^2<1$. 
For the cosmic monopole in the scalar field of amplitude $\eta$ one has the following correspondence:
 \begin{equation}
1- 8\pi G \eta^2=\alpha^2= \frac{c_\parallel^2} {c_\perp^2}\,.
\label{GlobalMonopole}
\end{equation}
The solid angle excess corresponds to the repulsive gravity, $G<0$, and super-Planckian scalar field, $\eta^2 >1/|G|$.

For $c_\parallel^2 =c_\perp^2\equiv c^2$ the metric is flat, $g^{ik}=c^2\delta^{ik}$,  at least far from the monopole. In cosmology this corresponds to the absence of the cosmic global monopole, or the absence of the scalar field in the vacuum, $\eta=0$. 
However, the planar phase monopole does not disappear: the  singularity remains in the vielbein field, while the metric has only the localized bump in the curvature and is flat (not conical)  at infinity. The tetrad field monopole in the 4D Euclidean space (torsional instanton) with the localized bump in the curvature and the flat metric at infinity was considered by Eguchi and Hanson\cite{EguchiHanson1978,EguchiHanson1979}, and Hanson and Regge.\cite{HansonRegge1979}

\section{Conclusion}

The planar phase provides an example, when the gravity for fermions and bosons can be essentially different.
While the fermions are described by the $4\times 5$ vielbein matrix $e^\mu_a$, the bosons are described by the conventional 4D metric 
$g_{\mu\nu}$. The vielbein with non-quadratic matrix $e^\mu_a$ may exist in other superfluid phases, including the ultracold fermionic gases.  In the presence of topological objects, they may give rise to exotic effective spaces and spacetimes, which are different for fermions and bosons. One may expect the similar effects in general relativity with degenerate metric.  Exotic monopole in gravity with degenerate tetrads was discussed for example in Ref.\cite{Sengupta2020}. It would be interesting to consider the transition from the planar phase to the $^3$He-B, where the Dirac fermions become massive.

  {\bf Acknowledgements}. This work has been supported by the European Research Council (ERC) under the European Union's Horizon 2020 research and innovation programme (Grant Agreement No. 694248).


\begin{thebibliography}{99}

\bibitem{Vollhardt1990} 
D. Vollhardt and P. Woelfe, 
{\it The Superfluid Phases of Helium 3}
(Taylor \& Francis, London, 1990).

\bibitem{Surovtsev2019} 
E.V. Surovtsev,
Phase diagram of superfluid $^3$He in a nematic aerogel in a strong magnetic field,
JETP {\bf 128}, 477--484 (2019).

\bibitem{Makhlin2014} 
   Y. Makhlin, M. Silaev, and G.E. Volovik,
Topology of the planar phase of superfluid $^3$He and bulk-boundary correspondence for
three-dimensional topological superconductors,
Phys. Rev. B {\bf 89} 174502 (2014).

\bibitem{Volovik2003} 
G.E. Volovik, 
{\it The Universe in a Helium Droplet}, 
Clarendon Press,  Oxford (2003).

\bibitem{Starobinskii1977}
D.D. Sokoloff, A.A. Starobinskii,	
On the structure of curvature tensor on conical singularities,
Dokl. Akad. Nauk SSSR {\bf 234}, 1043--1046 (1977),
Sov. Phys. Dokl. {\bf 22}, 312 (1977).

\bibitem{Barriola1989}
M. Barriola and A. Vilenkin,
Gravitational field of a global monopole,
Phys. Rev. Lett. {\bf 63}, 341-343 (1989).

\bibitem{Monopole1990}
D. Harari and C. Loustó,
Repulsive gravitational effects of global monopoles,
Phys. Rev. D {\bf 42}, 2626 (1990).

\bibitem{Monopole2001}
E.R. Bezerra de Mello,
Physics in the global monopole spacetime,
 Brazilian Journal of Physics  {\bf 31}, 211 (2001).
   
   
\bibitem{Bronnikov2002}
K. A. Bronnikov, B. E. Meierovichc and E. R. Podolyak,
Global monopole in general relativity,
JETP {\bf 95},  392--403 (2002).

\bibitem{Mavromatos2017}
N.E. Mavromatos and S. Sarkar,
Magnetic monopoles from global monopoles in the presence of a Kalb-Ramond field,
Phys. Rev. D {\bf 95}, 104025 (2017).

\bibitem{Petrov2019}
J. R. Nascimento, Gonzalo J. Olmo, P. J. Porfírio, A. Yu. Petrov and A. R. Soares,
Global monopole in Palatini $f({\cal R})$ gravity,
Phys. Rev.  D {\bf 99}, 064053 (2019).

\bibitem{Monopole2020}
E.A.F. Braganca, R.L.L. Vitoria, H. Belich, E.R. Bezerra de Mello,
Relativistic quantum oscillators in the global monopole spacetime,
Eur. Phys. J. C {\bf 80}, 206 (2020).

\bibitem{Volovik1998}
G.E. Volovik, 
Gravity of monopole and string and gravitational constant in $^3$He-A,   
Pisma ZhETF  {\bf 67}   666--671 (1998),
JETP Lett. {\bf 67}, 698--704  (1998); cond-mat/9804078. 

\bibitem{Nambu1977}
Y. Nambu, 
String-like configurations in the Weinberg-Salam theory,
Nucl. Phys. B {\bf 130}, 505 (1977).

\bibitem{Zhang2020}
G. E. Volovik, K. Zhang,
String monopoles, string walls, vortex-skyrmions and nexus objects in polar distorted B-phase of $^3$He,
Physical Review Research {\bf 2}, 023263 (2020),
arXiv:2002.07578.

\bibitem{EguchiHanson1978}
T. Eguchi and A.J. Hanson, 
Asymptotically flat self-dual solutions to euclidean gravity,
Phys. Lett. {\bf 74} B, 249 (1978).

\bibitem{EguchiHanson1979}
T. Eguchi and A.J. Hanson,
Self-dual solutions to euclidean gravity, 
Ann. Phys. {\bf 120}, 82--106 (1979).

\bibitem{HansonRegge1979}
A.J. Hanson and T. Regge, 
Torsion and quantum gravity, 
in: Proceedings of the Integrative Conference on Group Theory and Mathematical Physics, University of Texas at Austin, 1978,
Lecture Notes in Physics {\bf 94}, 354--361 (1979).


\bibitem{Sengupta2020}
S. Gera, S. Sengupta,
Emergent monopoles and magnetic charge,
arXiv:2004.13083.

\end{thebibliography}
\end{document}